\documentclass[twocolumn,aps,prd,superscriptaddress]{revtex4}
\usepackage{graphicx,rotating}
\usepackage{dcolumn}
\usepackage{epsfig}
\usepackage{multirow}
\usepackage{subfigure}

\input babarsym

\begin{document}
{\pagestyle{empty}



\title{
\Large{ \bf {\boldmath {Experimental Results on Flavour Oscillations and \CP-Violation}}}\\
\bigskip

\small{Prafulla Kumar Behera (on behalf of the \babar\ Collaboration)}. \\ 
\scriptsize{\bf The University of Iowa, Iowa City, USA. E-mail:behera@slac.stanford.edu}

\date{\today}
}






\begin{abstract}

30 years   ago M. Kobayashi and T. Masakawa  (won the Nobel Prize in physics 2008 for their famous theory),  gave a theory to explaining \CP\ violation.  This  theory predicted that large \CP\ violation would  be observed in \B\ mesons .  In the last 10 years, Both the \B-factories (\babar\ at Stanford Linear Accelerator Center, USA and Belle at KEK, Japan) have observed large \CP\ violation as predicted by the KM mechanism. The ground breaking experimental results of  Flavor  Oscillation and \CP\ violation in \B\ and $D$ mesons are  summarized and compared with the Standard Model expectations. 

\end{abstract}
\maketitle

}

\setcounter{footnote}{0}
\section{Introduction}
\label{sec:Introduction} 
  The origin of \CP\ violation in the Standard Model (SM) is contained in a complex phase within the Cabibbo-Kobayashi-Maskawa (CKM) matrix~\cite{CKM}. With one single phase, the SM predicts clear patterns for quark mixing and \CP\ violations, to be satisfied by all processes. The unitarity relation $V_{ud}V^{*}_{ub} + V_{cd}V^{*}_{cb} + V_{td}V^{*}_{tb} = 0 $ among the elements of the first and third columns of the CKM matrix is represented in the complex plane by a Unitarity Triangle (UT) with angles $\alpha = arg[-V_{td}V^{*}_{tb}/V_{ud}V^{*}_{ub}]$, $\beta = arg[-V_{cd}V^{*}_{cb}/V_{td}V^{*}_{tb}]$, $\gamma = arg[-V_{ud}V^{*}_{ub}/V_{cd}V^{*}_{cb}]$. The angle $\beta$, which is the phase of $V_{td}$ and appears in $\Bz-\Bzb$ mixing, induces mixing assisted \CP\ violation first observed by \babar\ and Belle in the \bpsiks\ decays in 2001.

  The \B-Factories have demonstrated that \CP\ violation in the \B meson system is consistent with the SM description given in Eq.[1]. 
\begin{equation}
 {A_{CP} (\delta t)} = S_{f}\textnormal{sin}({\delta m_{d} \delta t}) - C_{f}\textnormal{cos}({\delta m_{d}\delta t}).
\label{Eq1}
\end{equation}
 where $\delta t$ is the difference between the proper decay times of the two mesons and $\delta m_{d}$ is the mass difference of the \B\ meson mass eigenstates\footnote{Some authors, including the Belle collaboration use the symbols $\phi_{1}$, $\phi_{2}$, $\phi_{3}$ for the angles $\alpha$, $\beta$, $\gamma$, and $A_{f} = - C_{f}$ for the parameter describing direct \CP\ violation. In the present article we will follow the $\alpha$, $\beta$, $\gamma$, $C_{f}$ nomenclature.}. The sine term in Eq.~\ref{Eq1} comes from the interference between direct decay and decay after a $\Bz-\Bzb$ oscillation. A non-zero cosine term arises from the interference between decay amplitudes with different weak and strong phases (direct \CP\ violation) or from \CP\ violation in $\Bz-\Bzb$ mixing, where the latter is predicted to be small in the SM and has not been observed to date. The presence of new physics beyond SM could in general change this picture; for this reason it is very important to over-constrain the UT, making many independent measurements looking for inconsistencies in the SM. 
\section{Asymmetric Energy \ep\en Collision at \FourS}

\begin{figure}[t]
\begin{center}
\includegraphics[width=3.0in]{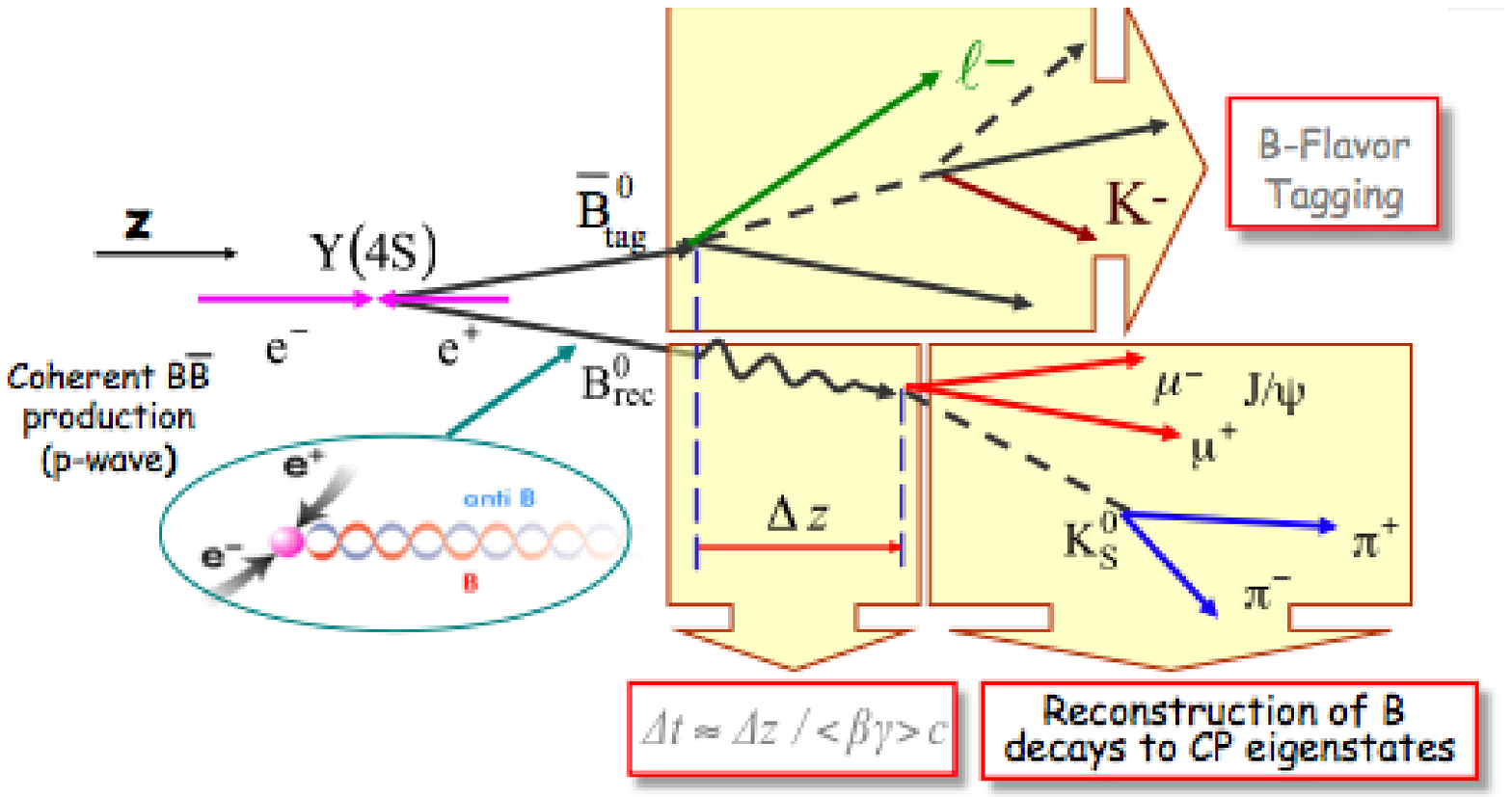}

\caption{ Asymmetric energy $\ep\en$ collision at $\FourS$.}
\label{asym}
\end{center}
\end{figure}

  Study of time-dependent \CP violation requires large data samples of \B\ meson decays with precise decay length measurements. The asymmetric energy $\ep\en$ colliders, PEP-II and KEKB, operating at \FourS\ facilitates this measurement by providing pair of $\B\Bbar$ from strong decays of the \FourS\ resonance in the $\ep\en$ rest system. Since the \B\ flight length in x-y is only $\sim 30 \mu$m as compared to $\sim$ 200 $\mu$m in z-direction, the approximation $\delta$r = $\delta$z = z$_{cp}$ - z$_{tag}$ can be made, where $z_{tag}$ and $z_{cp}$ are the z decay vertexes of the tagged \B\ with measured flavor and $z_{cp}$ is the CP-eigenstate vertex measured. Since flavor has to be conserved in the \BB\ system, at any given moment when the tagged \B\ is identified, the other \B\ has to be of the opposite flavor, as shown in Fig.~\ref{asym}, clocked as $\delta t$ = 0. Furthermore, because of the asymmetric nature, the separation distance between the two decay vertexes makes a good measurement of the decay distance possible; since the \B\ flight length in x-y is only $\sim 30 \mu$m as compared to $\sim$ 200 $\mu$m in z-direction, the approximation $\delta$r = $\delta$z = z$_{cp}$ - z$_{tag}$ can be made (Fig.~\ref{asym}). \babar\ and Belle collected a total of 531 \invfb, and $\sim$ 850 \invfb, respectively with \babar\ concluding its data taking in April 2008.

\section{Measurements of $\beta$ }
\subsection{ sin2$\beta$ from $\b \to \c\cbar\s$ }
 Interference between $\Bz \to f$ and $\Bz \to \Bzb \to f$ leads to time-dependent asymmetry Eq[1]. The $\jpsi\KS$ and $\b \to \c\cbar\s$ decays ($\jpsi\KL$, $\psitwos\KS$, $\chicone\KS$, $\etac\KS$, and \jpsi\Kstar) provide theoretically clean measurements of sin2$\beta$. They are dominated by a single decay amplitude and $S_{f}$ = sin2$\beta$ for $\jpsi\KS$. Extending to other $\b \to \c\cbar\s$ decays, we obtain sin2$\beta$ = -$\eta_{f} \times S_{f}$. The SM corrections to this relation are believed to be very small and \order($10^{-4}$)~\cite{Boos}.

\begin{figure}
    \centering

    {
        \includegraphics[width=1.4in]{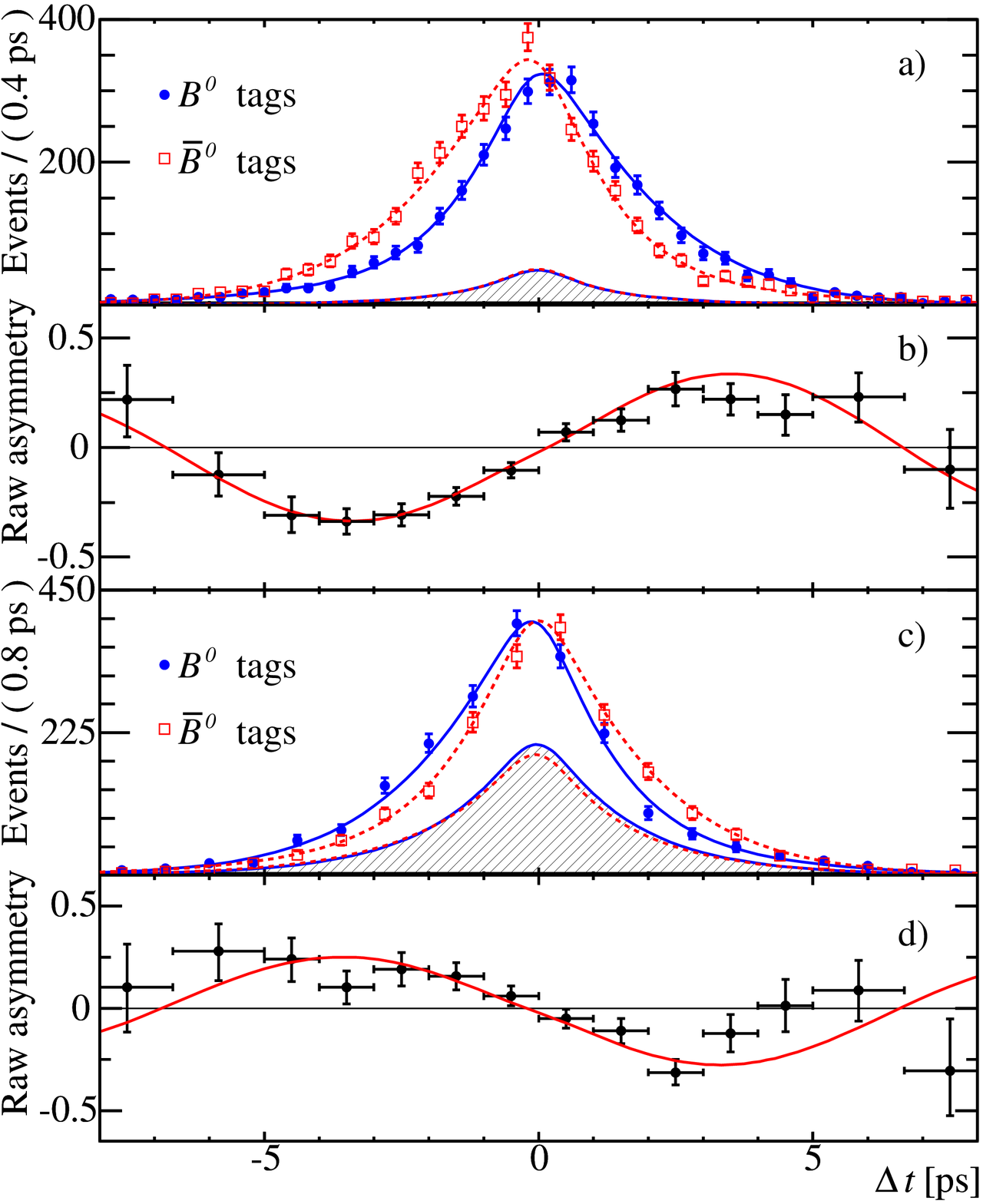}
    }

    {
        \includegraphics[width=1.7in]{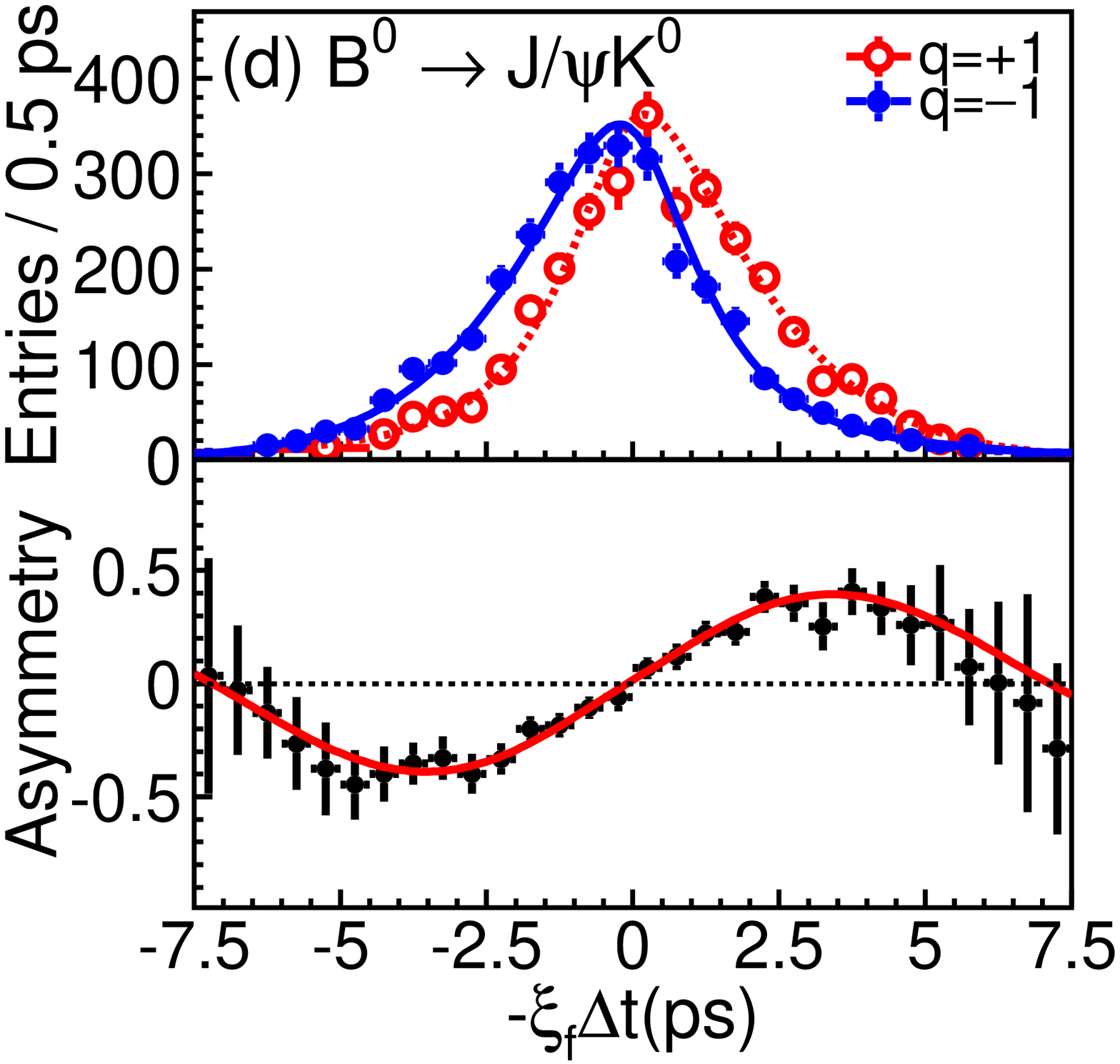}
    }
\caption{ Top figure shows the background subtracted $\Delta t$ distribution and the symmetries for events with good tags for \babar\ and the bottom figure for Belle ($\Bz\to\jpsi \KS$).}
\label{babar_cp}
\end{figure}





  The latest measurement from \babar~\cite{babarcp} are sin2$\beta$= 0.670 $\pm$ 0.031 $\pm$ 0.013, $C$ = 0.026 $\pm$ 0.020 $\pm$ 0.016\footnote{Here and in the following, unless otherwise noted, the first error is statistical and the second one systematic, they are combined if only one error is given}; the $\delta t$ distribution and the measured raw asymmetry are shown in Fig.~\ref{babar_cp}. Belle recently updated their results as shown in Fig.~\ref{babar_cp} and obtained sin2$\beta$ = 0.642 $\pm$ 0.031 $\pm$ 0.017, $C$ = 0.018 $\pm$ 0.021 $\pm$ 0.014~\cite{bellecp}. The new \babar\ and Belle average is sin2$\beta$ = 0.657 $\pm$ 0.025 using $\jpsi\Kz$ decay. Belle has updated their results for the $\psitwos\KS$ decay mode~\cite{bellepsi2s}, using 657 $\times~10^{6}$ \BB\ pairs, with sin2$\beta$ = 0.718 $\pm$ 0.090 $\pm$ 0.033 and $C~=~0.019~\pm~0.020~\pm~0.015$; both are in good agreement with the $\Bz \to \jpsi\KS$ measurement.    

\subsection{ Four-fold ambiguity of $\beta$}

  There remains a four-fold ambiguity for the value of $\beta$, $21.7^{0}$, $(21.7 + 180)^{0}$, $68.3^{0}$, and ($68.3 + 180)$. One approach for resolving this ambiguity is to measure cos2$\beta$, using a time-dependent angular analysis of $\Bz \to \jpsi \Kstarz (\KS\piz$) and $\Bz\to\Dz\piz$. \babar~\cite{babarpsikst} performed this analysis using using $88 \times$ $10^{6}$ \BB\ pairs and obtained $\textnormal{cos}2\beta = 2.72 ^{+0.05}_{-0.79}~\pm~0.27$ ($\textnormal{sin}2\beta = 0.731$). A similar analysis was performed by Belle~\cite{bellepsikst} using $275~\times~ 10^{6}$ \BB\ pairs obtaining $\textnormal{cos}2\beta = 0.87 \pm 0.74 \pm 0.12$ ($\textnormal{sin}2\beta = 0.726$), the error on measurement is too large to resolve the fourfold ambiguity. Belle's recent measurement of a time-dependent Dalitz analysis of $\Bz\to\Dz\piz$~\cite{belledpi0} disfavors the $68.3^{0}$ solution. 

\subsection{ sin2$\beta$ from $b \to c \cbar d$ decays}

 sin2$\beta$ can be potentially measured in  $\Bz \to \jpsi \piz$ and $\Bz \to D^{(*)+} D^{(*)-}$ decays. Main contribution from the decay $\Bz \to \jpsi \piz$ is from a color-suppressed internal spectator tree diagram while the $\Bz \to D^{(*)+} D^{(*)-}$ is dominated by color allowed tree diagram. The weak phase of the CKM matrix element involved is the same as in the $b \to \ccbar s$ decays; one would expect $C=0$ and $S = \textnormal{sin}2\beta$ in absence of penguin-mediated contributions in SM. The time dependent measurements of the decay $\jpsi\piz$ by Belle experiment are $S = -0.65 \pm 0.21 \pm 0.05$ and $C = -0.08 \pm 0.16 \pm 0.05$~\cite{bellejpsipiz}. The branching fraction of $\Bz \to \jpsi \piz$ can be used to constrain possible penguin contributions to the SU(3)-related decay \jpsi\KS~\cite{pierini}, and is measured to be $\BR({\jpsi\piz}) = (1.69 \pm 0.14 \pm 0.07) \times 10^{-5}$. Recently \babar\ measured $S = -1.23 \pm 0.21 \pm 0.04$ and $C = -0.20 \pm 0.19 \pm 0.03$ for \CP-even $\Bz \to \jpsi \piz$, which corresponds to a $4\sigma$ evidence for \CP violation~\cite{babarjpsipiz}. \babar's measurement is consistent with Belle's result.
  
  The decay $\Bz \to D^{(*)+} D^{(*)-}$ is a Vector-Vector (VV) final state. It can have contributions from  $L$ = 0, 1, 2 angular momentum states and therefore to both \CP-even and \CP-odd components. It is necessary to measure the \CP-odd fraction $R_{\perp}$, and then to take into account the dilution due to the admixture. Belle's preliminary result~\cite{bellepril} of $R_{\perp} = 0.116 \pm 0.042 \pm 0.004$ and \CP parameters: $S = -0.93 \pm 0.24 \pm 0.15$ and $C = -0.16 \pm 0.13 \pm 0.02$. The published \babar\ measurement~\cite{babardd} found a consistent value of $R_{\perp} = 0.143 \pm 0.034 \pm 0.008$ and \CP\ parameters: $S = -0.66 \pm 0.19 \pm 0.04$ and $C = -0.02 \pm 0.11 \pm 0.02$. Belle claims a $3.2 \sigma$ evidence of direct \CP violation in $\Bz \to D^{+}D^{-}$; $S = -1.13 \pm 0.37 \pm 0.09$ and $C = -0.91 \pm 0.23 \pm 0.06$~\cite{bellef}. This is unexpected in SM and not supported by \babar's measurement~\cite{babardpdm}. Both of the measurements from \babar\ are consistent with the SM prediction of tree dominance~\cite{Pham} and therefore with the result in $b\to c\cbar s$. Proposed new physics models could cause sizable corrections~\cite{YG}; it is therefore important to reduce experimental uncertainties further.

\section{ Measurement of $\alpha$ }
   The angle $\alpha$ is measured with a time-dependant analysis of decays of neutral \B\ mesons, $\Bz \to h^{+}h^{-}$, with $h$ = $\pi, \rho$, $a_{1}$. These are sensitive to an effective parameter $\alpha_{eff}$ due to the interplay of the Tree and the Penguin diagrams. Hence, One can determine $\alpha - \alpha_{eff}$~\cite{GL} using isospin relations in the decays  $\Bz \to h^{+}h^{-}, \Bz \to h^{0}h^{0},$ and $\Bpm \to h^{0}h^{\pm}$. The procedure of measuring the so-called ``isospin-triangles'' requires large dataset and leaves with up to eight-fold ambiguities. A less stringent relation sin$^{2}(\alpha - \alpha_{eff}) < \BR(\B \to h^{0}h^{0})/\BR(\B \to h^{0}h^{\pm})$ holds ~\cite{YGE}, which is more accessible with current datasets than the isospin analysis as it does not require to tag the flavor of the decaying \B.
\subsection{ $\alpha$ from $\B \to \pi\pi$ }

\begin{figure}
    \centering
    {
        \includegraphics[width=1.7in]{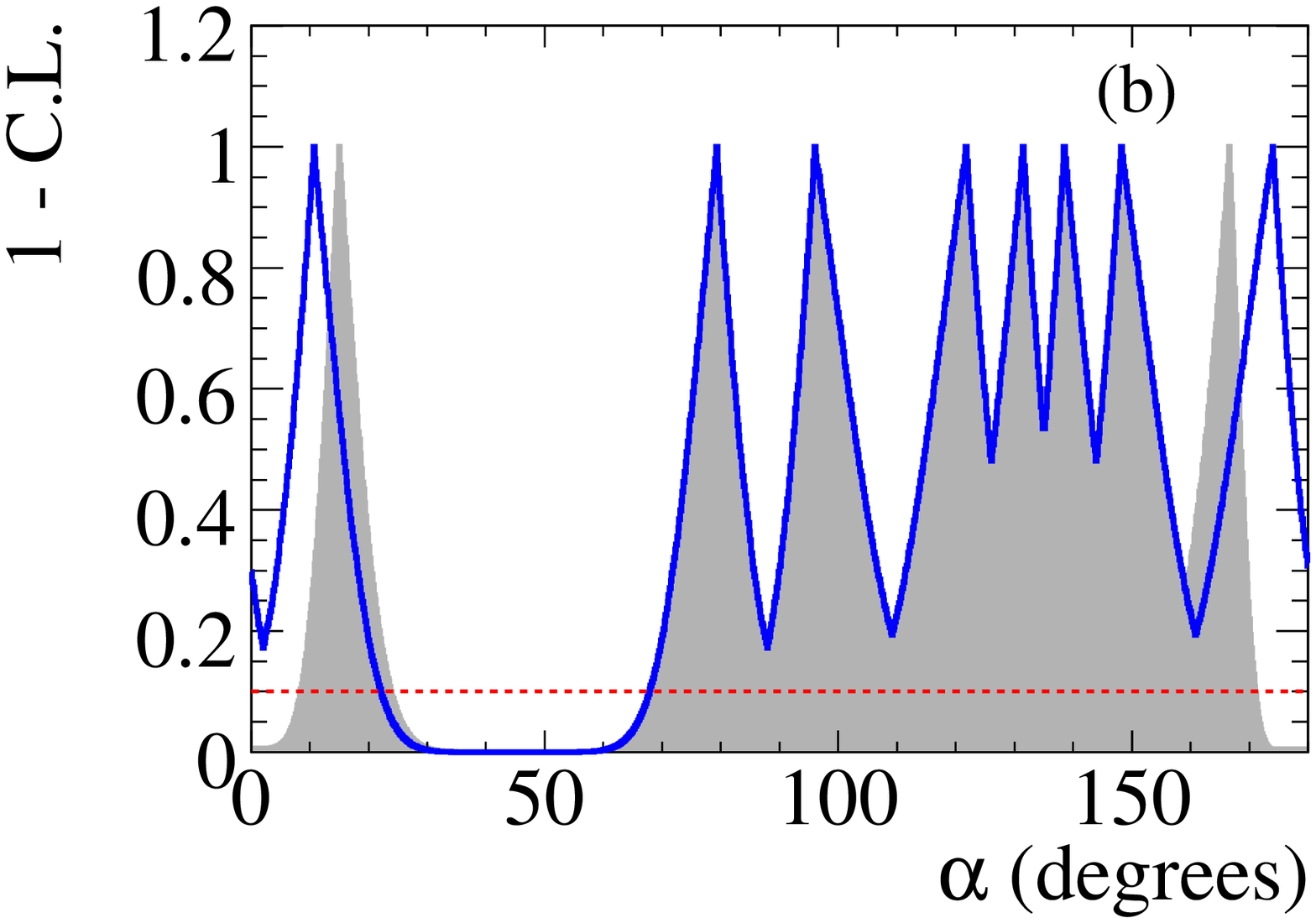}
    }
    {
        \includegraphics[width=1.4in]{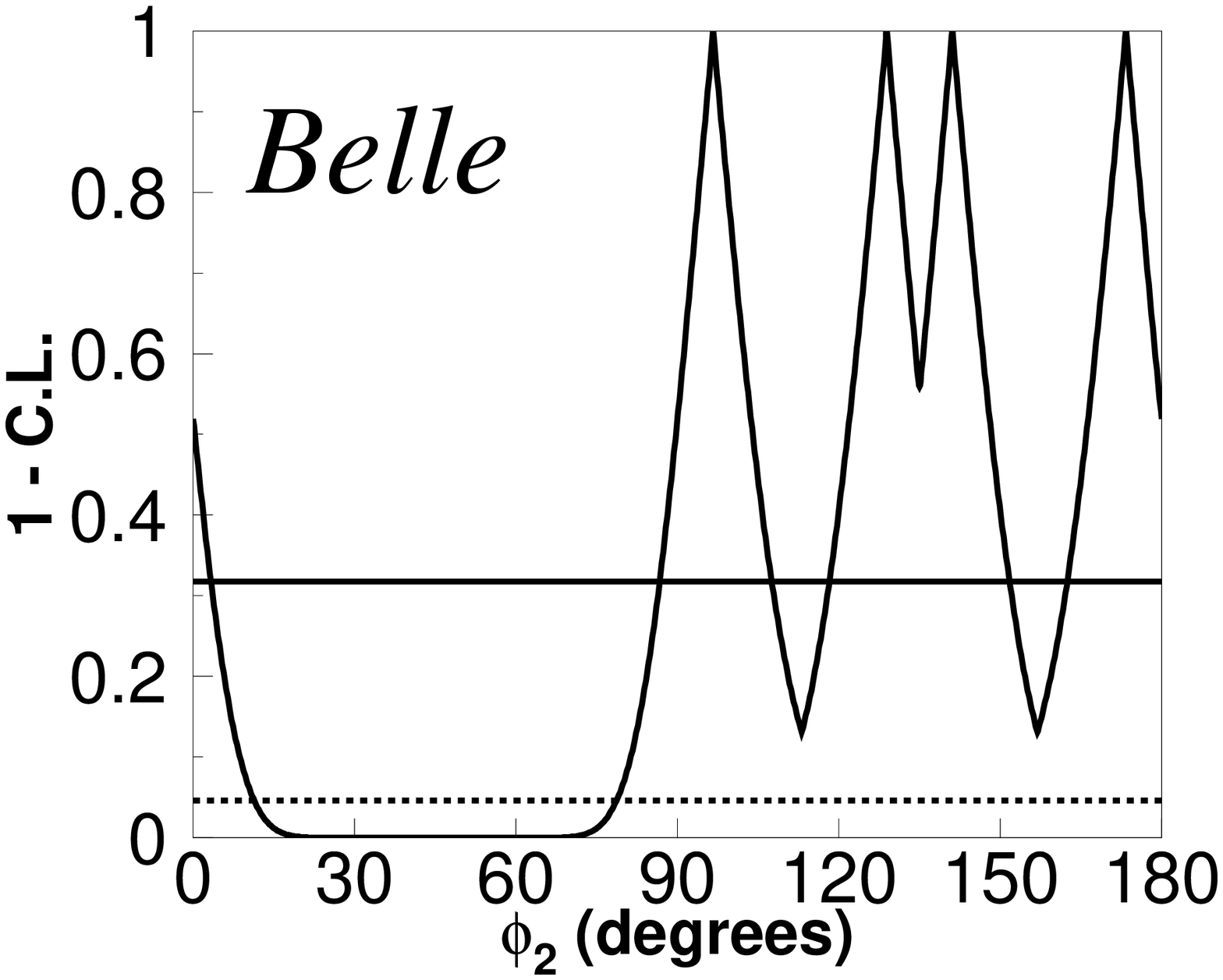}
    }
    \caption{Left Fig:$\alpha$ expressed as one minus the confidence level as a function of angle. An upper bound on $\delta \alpha$ of 39$^{0}$ at the $90\%$ CL. The curve shows the bounds on $\alpha$ using isospin method alone, while the shaded region shows the result with $SU(3)$ requirement. Right Fig: same values from Belle using isospin analysis and direct \CP asymmetry in $\Bz\to \pi^{0}\pi^{0}$ and $\BR{ B \to \pi\pi}$ decay. Solid line and dash line indicates C.L = $68.3\%$ and $95\%$, respectively.}
    \label{alphafig}
\end{figure}

   The decay $\B \to \pi\pi$ can be used to extract sin2$\alpha$.  \babar~\cite{babarpipi} and Belle~\cite{bellepipi} performed an isospin analysis to extract $\alpha$, using the available information ($S_{+-}, C_{+-}, C_{00}, B_{+-}, B_{+0}, B_{00}$), shown in Fig.~\ref{alphafig}. The preferred value of $\alpha = (96^{+10}_{-6})^{0}$ for \babar, which is consistent with Belle measurement of $\alpha = (97 \pm 11)^{0}$. 

\subsection{ $\alpha$ from $\B \to \rho\rho$}

  In $\Bz \to \rho^{+}\rho^{-}$ decays, a spin zero particle (the \Bz meson) decays into two spin 1 particles ($\rho^{\pm}$ mesons). Each $\rho^{\pm}$ meson decays into a $\pi^{\pm}\pi^{0}$ pair, where the quark content of the final decay product is similar to $\pi\pi$. The presence of a $\piz$ in the final state adds to experimental challenge of the measurement. The \CP analysis of $\Bz \to \rho^{+}\rho^{-}$ is further complicated by the presence of one amplitude with longitudinal polarization and two amplitudes with transverse polarization. The decay is observed to be dominated by the longitudinal polarization. There is good agreement between the measurements in $\rho^{+}\rho^{-}$ from \babar~\cite{babarrr} and Belle~\cite{bellerr} for \CP\ violation. The HFAG average for the longitudinal components are $C_{\rho^{+}\rho^{-}} = -0.06 \pm 0.13$, $S_{\rho^{+}\rho^{-}} = -0.05 \pm 0.17$. For the first time, \babar~\cite{babarrprm} showed a preliminary time-dependant measurement of $\Bz \to \rho^{0}\rho^{0}$ decay. Using $427 \times 10^{6}$ \BB\ pairs, \babar\ measures $\BR(\Bz \to \rho^{0}\rho^{0}) = (0.84 \pm 0.29 \pm 0.17) \times 10^{-6}$, $f_{L} = 0.70 \pm 0.14 \pm 0.05$, $S_{L} = 0.5 \pm 0.9 \pm 0.2$, $C_{L} = 0.4 \pm 0.9 \pm 0.2$. Consistently, Belle~\cite{bellerprm} set an upper limit of $\BR(\Bz \to \rho^{0}\rho^{0}) < 1.0 \times 10^{-6}$ at a $90\%$ confidence level (C.L).

\subsection{ $\alpha$ from $\B \to \rho\pi$}

   The decay $\Bz \to \pip\pim\piz$ can be used to measure the angle $\alpha$, in principle however this is not a \CP\ eigenstate decay. Performing an isospin analysis is extremely complicated involving pentagonal relationship among different amplitudes. It can not be solved for the 12 unknowns with present statistics. It was, however, pointed out~\cite{QS} that the variation of the strong phase of the interfering $\rho$ resonance in the dalitz plot provides the necessary degrees of freedom to constrain $\alpha$ with only the irreducible ($\alpha \to \alpha + \pi$) ambiguity. The \babar\ experiment thus performed this analysis and measured $\alpha = (87^{+45}_{-13})^{0}$~\cite{babarrp}. Belle~\cite{bellerp} performed this analysis resulting in a tighter constraint of $68^{0} < \alpha < 95^{0}$ at $95\%$ C.L for the solution compatible with SM.   
  
\subsection{ $\alpha$ from $\B \to a_{1}\pi$}

  The analysis of the decay $\Bz \to a_{1}^{\pm}\pi^{\mp}$ where $a_{1}^{\pm} \to \pi^{\mp}\pi^{\pm}\pi^{\pm}$ has been used to extract $\alpha$ with a quasi-two-body approximation.  \babar~\cite{babara1p} performed this analysis using 349 $\fb^{-1}$ of data and measured $\alpha_{eff}$ = $(78.6 \pm 7.3 )^{0}$, with no evidence for the direct or mixing induced \CP\ violation. Once the measurements of branching fractions for SU(3)-related decays ($\Bp \to a_{1}^{+}(1270)\pi$  or $\Bp \to a_{1}^{-}(1260)\Kp$ and the decays $B \to K_{1}(1270)\pi$ and $B\to K_{1}(1400)\pi$) become available, quantitative bounds on $\delta \alpha$ obtained with method~\cite{MGZ} will provide significant constraints on the angle $\alpha$ through the measurement of $\alpha_{eff}$ in $\Bz\to a_{1}^{\pm}(1260)\pi^{\mp}$ decays. Recently \babar~\cite{babara1k} performed an analysis to measure these decay modes to constrain $|\alpha - \alpha_{eff}|$. They observed the charged and neutral decay modes for the first time and found no evidence for direct \CP asymmetry.

\section{ Measurement of $\gamma$ }
 At the beginning of the \B-Factory $(\ep\en\to\FourS\to\BB)$ era, it was thought to be not possible to measure $\gamma$ and it could only be measured at hadron collider like CDF and D0. Latter several methods have been pursued using different common decay modes such as \CP\ eigenstates (Gronau, London, and Wyler (GLWmethod)~\cite{MGDL}), suppressed $K\pi$ charge combinations (Atwood, Dunietz, and Soni (ADS method)~\cite{DISoni}), and $\KS\pip\pim$ decays (Giri, Grossman, Soffer, and Zuppan (GGSZ method)~\cite{Giri}).
\subsection{ The GLW method }
In the GLW method, \CP\ eigenstates with even or odd parity $D_{cp}^{\pm}$ are obtained final states common to \Dz\ and \Dzb. The branching fractions $\BR^{\pm}_{{\CP}^{\pm}}$, $\BR^{-}_{\Dz}$, and $\BR^{+}_{\Dzb}$ have been measured\footnote{The superscripts refer to the charge of the \B, the subscript to the \CP\ parity of the $D$ eigenstate $D_{\CP}^{\pm}$ or to the flavor of te $D$ for flavor-specific decays respectively.}, where the observables are constructed from their ratios, such as $R_{\CP^{\pm}} \equiv (\BR^{-}_{{\CP}^{\pm}} + \BR^{+}_{\CP^{\pm}}) / (\BR^{-}_{\Dz K} + \BR^{+}_{\Dzb K})/2$ and $A_{\CP}^{\pm} \equiv (\BR^{-}_{{\CP}^{\pm}} - \BR^{+}_{{\CP}^{\pm}})/ (\BR^{-}_{{\CP}^{\pm}} + \BR^{+}_{{\CP}^{\pm}})$. \babar~\cite{babarglw} published a measurement of $\B^{\pm} \to \Dstar\Kpm$, with decay modes $\Dstarz \to \Dz\g, \Dz\piz,$ and $\Dz(\Dzb)$ reconstructed in \CP-even ( $\Kp\Km, \pip\pim$), \CP-odd ($\KS\piz, \KS\omega, \KS\phi$), and flavor-specific modes ($\Kp\pip$). They obtained $A_{\CP^{+}} = -0.11 \pm 0.09 \pm 0.01$, $A_{\CP^{-}} = +0.06 \pm 0.10 \pm 0.02$, $R_{\CP^{+}} = 1.31 \pm 0.13 \pm 0.04$, $R_{\CP^{-}} = 1.10 \pm 0.12 \pm 0.04$. The accuracy of this measurement does not allow a determination of $\gamma$ with GLW method alone but it contributes to improving the overall precision when combined with other methods.

\subsection{ The ADS method }
  
   Decays with similar overall amplitudes are selected to maximize the interference and therefore sensitivity to \CP\ asymmetries. Favored $\B \to D$ decays followed by suppressed $D$ decays, or vice-versa provide an ideal data set for these studies. It is possible to define a ratio of branching fractions of suppressed and favored decays as $R_{ADS} = \BR_{B\to D_{sup}K}/\BR_{B\to D_{fav}K}$ = $r^{2}_{D} + r^{2}_{B} + 2 r_{D} r_{B}$cos$\gamma$cos$\delta_{B}$, and \CP\ asymmetries as $A_{ADS} = (\BR_{\B^{-}\to D_{sup}K^{-}} - \BR_{B^{+} \to D_{sup}K^{+}})/ (\BR_{\B^{-}\to D_{sup}K^{-}} + \BR_{B^{+} \to D_{sup}K^{+}}) = 2 r_{D} r_{B} \textnormal{sin} \gamma \textnormal{sin}\delta_{B}/R_{ADS}$. Belle~\cite{belleads} recently published measurements of the decay chain $\Bm \to D\Km$, $D \to \Kp\pim$, based on 657 $\times 10^{6}$ \BB\ pairs. They do not observe a statistically significant signal in the suppressed mode, obtaining $R_{ADS} = ( 8.0^{+6.3}_{-5.7} {^{+2.0}_{-2.8}}) \times 10^{-3}$, and $A_{ADS} = ( -0.13^{+0.97}_{-0.88}\pm 0.26)$. These results are used to set a $90\%$ C.L upper limit of $r_{B} < 0.19$.

\begin{figure}[t]
    \centering
    {
        \includegraphics[width=1.5in]{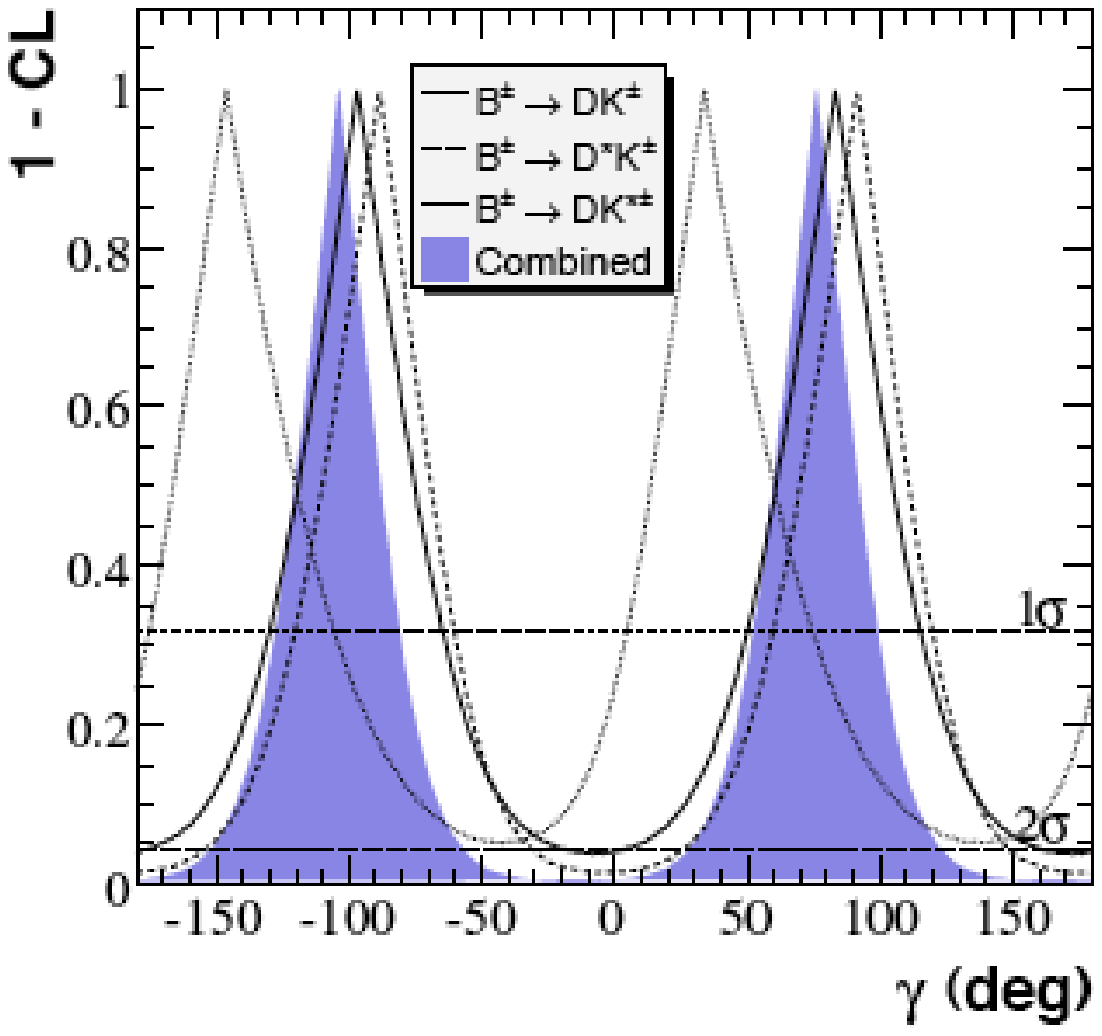}
    }
    {
        \includegraphics[width=1.5in]{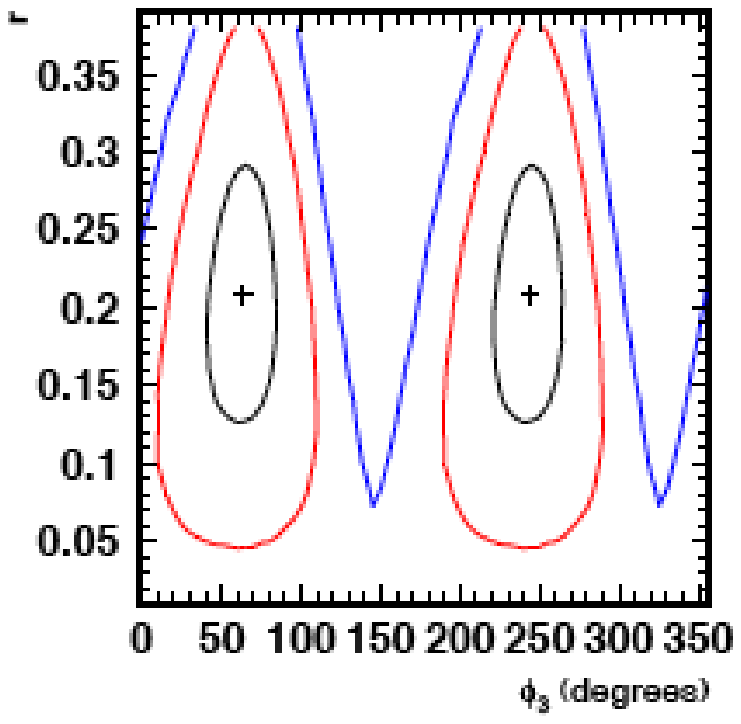}
    }
    \caption{ Left Fig: For \babar, 1$-$ CL as a function of $\gamma$ including statistical and
systematic uncertainties and their correlations. The dashed (upper) and dotted (lower) horizontal lines correspond to the one- and two-standard deviation intervals, respectively. Right Fig: For Belle, Projections of confidence regions for $\Bp \to \Dstar\Kp$ mode onto the (r, $\gamma$) planes. Contours indicate projections of one, two and three standard deviation regions.}
\label{gamma}
\end{figure}
  
\subsection{ GGSZ method }

  The three body decays of $\Dz$ and $\Dzb$ such as $\KS\pip\pim$ or $\KS\Kp\Km$ depend on the interference of Cabibbo allowed, doubly-Cabibbo suppressed, and \CP\ eigenstate, decay amplitudes. The $\Bpm \to \Dz (\Dzb) \Kpm$ amplitude can be written as $\calA_{\mp}^{(*)}(m_{-}^{2},m_{+}^{2}) \propto \calA_{\Dmp} + \lambda r_{B}^{*}e^{i(\delta_{B}^{*}\pm \gamma)} \calA_{\Dpm}$, where $m_{-}^{2}$ and $m_{+}^{2}$ are the squared invariant masses of the $\KS\pim$ and $\KS\pip$ for $\lambda = -1 $ for $\Dstar \to \Dz \gamma$ and +1 for other decays, respectively and $\calA_{\Dp}(\calA_{\Dm})$ are the amplitudes of the $\Dz(\Dzb) \to \KS\pip\pim$ decay. These decays are described in a detailed model involving several intermediate resonances extracted from large control samples of flavor-tagged $\Dstarp \to \Dz \pip$ decays produced in $c\cbar$ events. The Cartesian variables $x_{\mp}^{*} = r_{B}^{*}$ cos($\delta_{B}^{*} \mp \gamma$), $y_{\mp}^{*} = r_{B}^{*}$ sin($\delta_{B}^{*} \mp \gamma$) are used by the experiments to avoid the bias $r_{B}^{*}$ being positive definite. Belle reconstructed the decays $\Bmp \to \Dstarz\Kmp$, with $\Dstarz \to \Dz\piz$ and $\Dz \to \KS\pip\pim$, and measured $r_{B} = 0.16 \pm 0.04$, $r_{B}^{*} = 0.21 \pm 0.08$ and $\gamma = (76^{+12}_{-13})^{0}$~\cite{bellekspp} using $657 \times 10^{6}$ \BB\ pairs. \babar\ recently published a result based on $383 \times 10^{6}$ \BB\ pairs~\cite{babarkspp}. In addition to the mode used by Belle, \babar\ also reconstructed the decays $\Dstarz \to \Dz\gamma$, $\Bpm\to \Dz\Kstarpm[\KS\pipm]$, and $\Dz \to \KS\Kp\Km$. \babar\ determines the $(x,y)_{\mp}^{*}$ parameters with the same accuracy as Belle. \babar\ data favors smaller $r_{B}$ values ($r_{B} = 0.086 \pm 0.035, r_{B}^{*} = 0.135 \pm 0.51, kr_{S}= 0.163^{+0.088}_{-0.105})$\footnote{ The amplitude ratio $\Bpm \to \Dz \Kpm$ events is described by $kr_{S}$, with $k$ taking in to account non-resonant $\KS\pi^{\pm}$ contributions.} , and thus a larger error on $\gamma$ ( $\gamma = (76^{+23}_{-24})^{0}$. The result for $(x,y)_{\mp}$ in the $\Bmp \to \Dz\Kmp$ decay mode from \babar\ and Belle are shown in Fig.~\ref{gamma}. 

\section{ sin$2\beta$ from $b \to q\qbar s $ Penguin Decays }
   The decay $\Bz \to \phi\KS$ is dominated by a penguin diagram. The SM amplitude is expressed by 
\begin{eqnarray}
 P & \sim &  V_{ub}^{*}V_{us}P^{u} + V_{cb}^{*}V_{cs}P^{c} + V_{tb}^{*}V_{ts}P^{t} \\ 
   & \sim & V_{cb}^{*}V_{cs}(P^{c} - P^{t}) + V_{ub}^{*}V_{us}(P^{u}-P^{t}) 
\end{eqnarray}
  Where $V_{tb}^{*}V_{ts}$ can be eliminated using one of the CKM unitarity relations, $V_{cb}^{*}V_{cs} + V_{ub}^{*}V_{us} + V_{tb}^{*}V_{ts}$ = 0. The first term, $\order{(A\lambda^{2})}$, contains CKM elements which are identical to $\b \to c\cbar s$ and leads to sin2$\beta$. The second term, $\order{(A\lambda^{4}(\rho - i\eta))}$ is suppressed compared to the first term.
    
   For other $\b \to s$ penguin modes, there are additional small contributions. For $\Bz \to \eta^{\prime} \KS, f^{0}\KS$, a contribution from $b \to u $ tree diagram, $\order(A\lambda^{4}(\rho - i\eta))$, can be present. The $\Bz \to \piz \KS$ and $\omega\KS$ can also have a contribution from the $\b \to u $ tree diagram. In addition, these modes contain $b \to s d\dbar$ instead of $b \to ss\sbar$. Considering these effects, the SM corrections up to $\order(\lambda^{2}) \equiv 5\%$  can be possible in the extraction of sin2$\beta$ and the magnitudes of corrections can vary in different modes~\cite{GLN}. A larger deviation exceeding these corections will be an indication of new physics in penguin loops.

\begin{figure}[t]
\begin{center}
\includegraphics[width=2.7in]{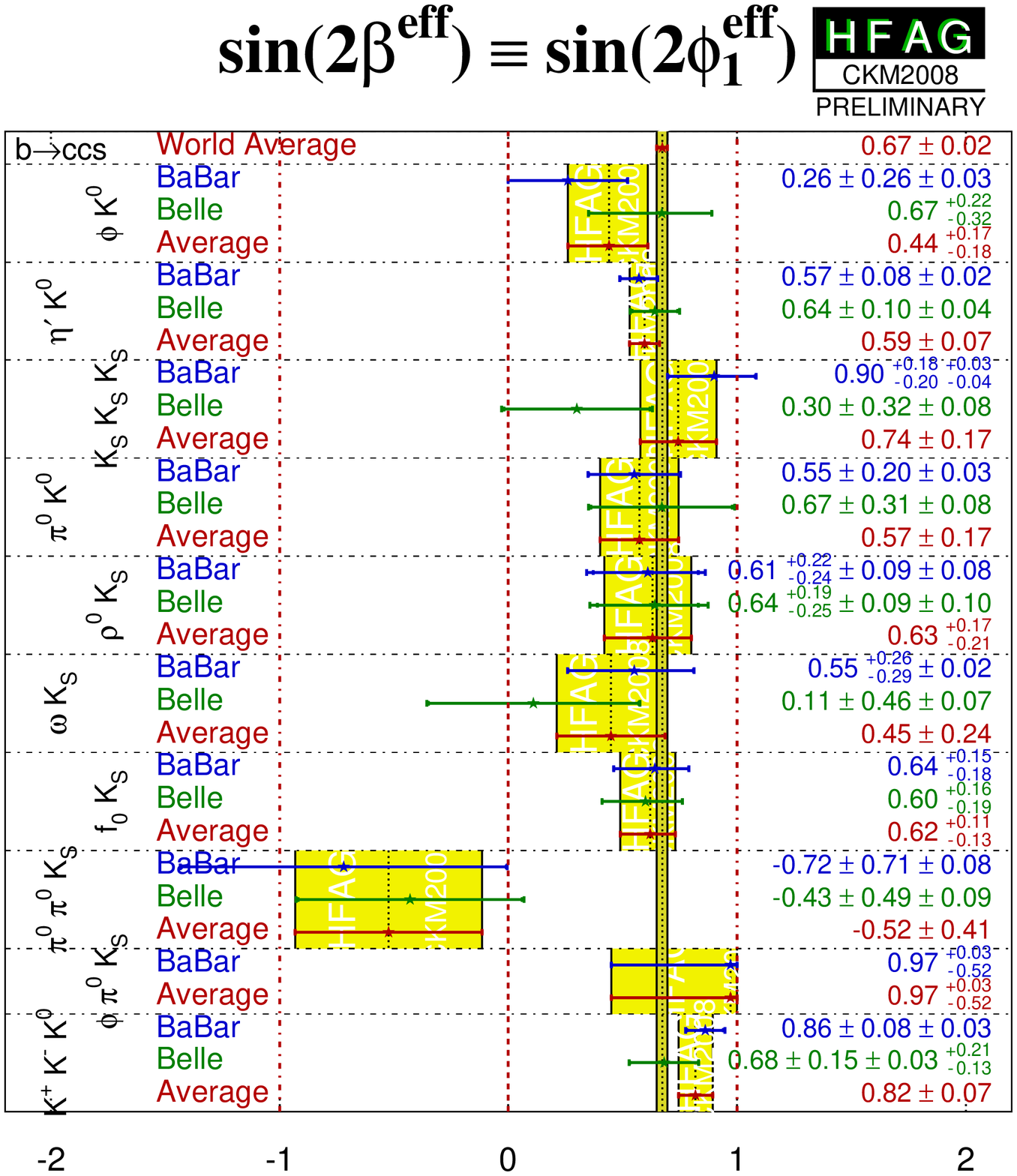}
\caption{ Summary of effective sin2$\beta$ measurement in $b\to s $ decay modes, compared to the world average sin2$\beta$ value in $\b \to \c \cbar s$. }
\label{HFAG_BSP}
\end{center}
\end{figure}

  Individual results from \babar\ and Belle, as well as their averages for each decay mode of $b\to s $ penguin and most recent sin2$\beta$ from $b \to c\cbar s$~\cite{HFAG}, shown in Fig.~\ref{HFAG_BSP} are mutually consistent. All sin2$\beta_{eff}$ measurements, except $\phi\KS$, $\piz\piz\KS$, and $\Kp\Km\Kz$ are within $\sim$ 1$\sigma$ from the value of sin2$\beta$ = 0.67 $\pm$ 0.02 from $b \to c\cbar s$.
\section{ \CP\ violation in \BB\ Mixing}

  The SM allows \CP\ violation in \BB\ mixing itself in analogy to $\epsilon_{k}$ in \Kz\ system~\cite{PDG}. This leads to $|q/p| \neq~1$, where $p$ and $q$ are the coefficients relating the mass and flavor eigenstates of the neutral \B\ mesons, because $M_{12}$ and $\Gamma_{12}$ have different phases. In SM we expect the observable effect to be $1- |\frac{q}{p}|^{2} \simeq Im(\frac{\Gamma_{12}}{M_{12}}) \sim \order(10^{-3}).$
      Observation of a significantly large effect would be an exciting result. \babar~\cite{babarmix} has measured $|\frac{q}{p}| - 1 = (-0.8 \pm 2.7 \pm 1.9 ) \times 10^{-3}$ using  $232 \times 10^{6}$ \BB\ pairs. Belle~\cite{bellemix} measured $|\frac{q}{p}| = 1.0005 \pm 0.0040 \pm 0.0043$. The results from \babar\ and Belle are consistent with unity. This implies that \CP\ violation in \Bz-\Bzb mixing is below the $\order(10^{-2})$ level. These measurements are already limited by their systematic uncertainties.  
\begin{figure}[t]
\begin{center}
\includegraphics[width=2.5in]{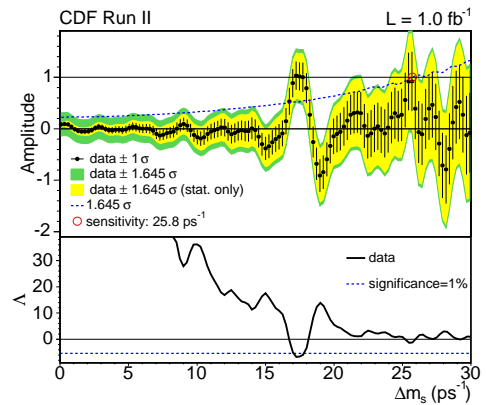}
\caption{ Upper figure shows the measured amplitude values and uncertainties versus the \Bs-\Bsb oscillation frequency $\Delta m_{s}$ for all decay modes combined. Lower plot: The logarithm oscillation. }
\label{HFAG_CDF}
\end{center}
\end{figure}
\section{ Mixing and \CP\ violation in  the \Bs system  }

  The determination of \Bs - \Bsb oscillation frequency related to $\Delta m_{s}$ can be used to extract the magnitude of $V_{ts}$, one of the nine elements of the CKM matrix. The CDF experiment~\cite{CDF} reported an observation of \Bs - \Bsb mixing, measuring $\delta m_{s} = 17.77 \pm 0.10 \pm 0.07$ ps$^{-1}$ with a data sample of 1 fb$^{-1}$ from $p\bar{p}$ collisions at $\sqrt{s}$ = 1.96 TeV. The amplitude values and $\Bs-\Bsb$ oscillation frequency $\Delta m_{s}$ are shown in Fig.~\ref{HFAG_CDF}, and it is consistent with the SM expectation. Using this value of $\Delta m_{s}$ they derived $|V_{td}/V_{ts}| = 0.2060 \pm 0.0007$ (exp)$^{+0.0081}_{-0.0060}$(theor), and found no evidence of \CP\ violation. The D0~\cite{d0bs} experiment also measured a stringent limit of $17 < \delta m_{s} < 21$ ps$^{-1}$ at $90\%$ C.L with $\sim$ 1 fb$^{-1}$. 
\section{ Mixing and \CP\ violation in  the \Dz system  }

 The \Dz\ system is unique among the neutral mesons because it is the only one in which mixing proceeds via intermediate states with down-type quarks (up-type quarks in supersymetric models). The \Dz-\Dzb rate is expected to be small ($10^{-4}$ or less) in SM. \babar~\cite{babarddm} found evidence of \Dz-\Dzb mixing in the variation in decay time distribution for \Dz\to \Kp\pim compared to that of the Cabibbo-favored decay \Dz\to\Km\pip.   Belle~\cite{belledd} also found evidence of \Dz-\Dzb mixing with $3.2\sigma$ using 540 $\fb^{-1}$ data. More recently the CDF~\cite{CDFdd} has found evidence for mixing in \Dz\ decays. No evidence for \CP\ violation is found by \babar, Belle, and CDF experiments.

\section{ Summary }

  The \B-Factories have observed \CP\ violation in several \B\ decays. All categories of \CP\ violation in \B\ decays are consistent with SM expectations. The precision of sin2$\beta$ measurements is now measured to better than $4\%$ and significantly constrains the UT. The decay $\B \to \pi\pi, \rho\rho, \rho\pi, $ and $a_{1}\pi$ are used to study $\alpha$. The decay $\rho^{+}\rho^{-}$ gave stringent constraint on $\alpha$. There are other decays which are being investigated and hopefully will contribute to the measurement of $\alpha$. At the begining of the \B-Factory ($\ep \en \to \FourS \to \BB$) era, it was thought to be not possible to measure the angle $\gamma$. The dalitz analysis of $\Bpm\to\Dz\Kpm$ has the best sensitivity for $\gamma$. The decay $b\to s $ penguin is of prime focus because of sensitivity to physics beyond SM.

    Mixing but no evidence of \CP\ violation has been observed in \Bz-\Bzb, \Bs-\Bsb, and \Dz-\Dzb. Constraints on both $\Delta\Gamma$ and $\Delta m $ for \Dz-\Dzb mixing will allow further studies of \CP violation which remains a potential ground for observing New Physics.

\end{document}